\documentstyle[preprint,aps]{revtex}
\renewcommand{\baselinestretch}{1}
\begin{document}
\title{Adiabatic Berry Phase and Hannay Angle for Open Paths}
\author{Arun Kumar Pati}
\address{Theoretical Physics Division, 5th Floor, Central Complex}
\address{Bhabha Atomic Research Centre, Bombay-400 085, INDIA}
\date{\today}
\maketitle
\begin{abstract}
 We  obtain  the  adiabatic Berry phase by defining a generalised
gauge potential whose line integral gives the phase holonomy  for
arbitrary  evolutions  of  parameters.  Keeping  in mind that for
classical  integrable  systems  it  is hardly clear how to obtain
open-path Hannay angle, we establish  a  connection  between  the
open-path  Berry phase and Hannay angle by using the parametrised
coherent state approach. Using the semiclassical wavefunction  we
analyse the open-path Berry phase and obtain the open-path Hannay
angle.  Further, by expressing the adiabatic Berry phase in terms
of   the   commutator   of   instantaneous  projectors  with  its
differential and using  Wigner  representation  of  operators  we
obtain  the Poisson bracket between distribution function and its
differential. This enables us to talk about the  classical  limit
of  the  phase  holonomy  which  yields  the  angle  holonomy for
open-paths. An operational definition of Hannay angle is provided
based on the idea of classical limit of quantum mechanical  inner
product.  A probable application of the open-path Berry phase and
Hannay angle to wave-packet revival  phenomena  is  also  pointed
out.

\end{abstract}

\newpage

PRPOSED  RUNNING  HEAD:  {\sl Adiabatic Berry  phase  and  Hannay  angle...}\\

NAME OF AUTHOR: Arun Kumar Pati.\\

MAILING ADDRESS:\\

Dr. Arun Kumar Pati.\\
Theoretical Physics Division\\
5th Floor, Central Complex\\
Bhabha Atomic Research Center\\
Mumbai (Bombay)-400 085, INDIA.\\

TELEPHONE: 91-22- 551 9946\\
FAX:       91-22- 556 0750/ 556 0530\\

EMAIL: krsrini@magnum.barc.ernet.in\\

\newpage

\centerline{1. {\bf Introduction}}

\par
     In  recent  years,  the  quantal phase holonomy \cite{bs} of
purely geometrical origin has played an important and fundamental
role in diverse areas of physics. Berry \cite{mb} discovered this
in  quantum  adiabatic  context,  where  the  quantal  eigenstate
acquires  an  extra  phase  when the Hamiltonian of the system is
adiabatically transported arround  a  closed  path  in  parameter
space.  At classical level there is a similar effect, namely, the
angle holonomy, discovered by Hannay  \cite{jh}.  For  integrable
systems  (where  it  is  possible to write the Hamiltonian of the
system in terms of action and angle variables), Hannay  angle  is
nothing but an extra angle shift picked up by the angle variables
of  the  classical  system  when the parameters undergo adiabatic
change along a closed path in  the  parameter  space.  After  the
importance  of  Berry's  discovery  was realised in many areas of
physics, it was liberated from  its  restrictions  to  adiabatic,
periodic  variations  of  Hamiltonian  evolutions.  Aharonov  and
Anandan \cite{yj} showed the existence of  the  geometric  phases
for  non-adiabatic,  cyclic  evolutions of quantal wavefunctions.
Samuel and Bhandari  \cite{sb}  generalised  the  idea  of  phase
holonomy   for  non-cyclic,  non-unitary  evolutions  of  quantum
systems. Mukunda and Simon \cite{ms} have generalised the concept
of geometric phase using kinematic concepts  of  the  ray  space.
Recently,  the  present author generalised it further to the case
of non-cyclic, non-unitary and non-Schr{\"o}dinger evolutions  of
the   quantum   systems   \cite{ak}.   Notwithstanding  the  wide
generalisation of the Berry phase, its classical counterpart  the
Hannay   angle  has  not  been  generalised  further  except  for
non-adibatic cases. Berry  and  Hannay  \cite{bh}  have  obtained
classical  non-adiabatic  angle  as the holonomy of a non-trivial
connection in the phase-space bundle. The Hannay angle  can  also
be  understood as an angle shift in transporting a classical tori
in phase space \cite{an}. Therefore, any  attempt  to  generalise
and  understand  the  classical  angle holonomy for open paths is
quite challenging.

\par
In  this paper we generalise the Berry phase and Hannay angle for
an adiabatically evolving system with non-cyclic variation of the
external paramaters of the Hamiltonian. Before achieving that  we
provide  a  gauge  potential  description  of the open-path Berry
phase. This defines a quantum one-form whose line integral  gives
the  Berry  phase  during  an  arbitrary  variations  of external
parameters. Using the parametrised  coherent  state  approach  we
establish  a  connection  between  the  Berry phase and open-path
Hannay angle. Also, we obtain the open-path Berry  phase  in  the
semiclassical  limit  and  relate it to Hannay angle. Further, we
express the quantum one-form in terms of instantaneous projection
operators and study its classical limit using the  correspondence
between the quantum commutator and Poisson bracket. Here, we have
used the Wigner representation of quantum mechnaical distribution
function  and  phase space functions. The generalisation of Hanny
angle will have many important applications such as  wave  packet
revivals  \cite{cj},  field  theoretical models with fermions and
Grasmannian systems \cite{gt}. The present work will be  a  first
step in this direction. We will not give a treatment of open path
Hannay  angle  based  on classical Hamiltonian and its cannonical
transformation to action- angle variables, rather we will  define
the  adiabatic  Berry phase for open paths in parameter space and
obtain the Hannay angle as a semiclassical limit of  the  former.
For  arriving  at  Hannay  angle  the  following  result  will be
invoked: The connection between the Hannay angle and Berry  phase
\cite{mv}  is  valid not only for the adiabatic closed-excursions
but also for the open-excursions  in  the  parameter  space.  The
reason  for doing this is that there is a difficulty in attacking
the problem purely at classical level.  For  integrable,  bounded
motions  of classical systems action variables are the classical,
adiabatic invariants (in quantum case, the quantum number  is  an
adiabatic invariant). These angle variables have some unavoidable
arbitrariness  in  their  definition and they can not be compared
belonging to distinct initial and final  Hamiltonians  \cite{jh}.
They  can  be  compared,  however,  if  the Hamiltonian is varied
arround a closed loop in parameter space so that the initial  and
final  Hamiltonians  are same. Then one can make the Hannay angle
coordinate independent (in quantum case  this  is  equivalent  to
making the Berry phase gauge invariant). If we wish to define the
Hannay  angle  for  open  paths  from classical considerations we
would face the problem of comparing the angle variables belonging
to distinct initial and final Hamiltonians. However, at a quantum
mechanical level there is no problem of comparing the  phases  of
two  distinct (they do not form same equivalence classes) initial
and final non-orthogonal vectors. Therefore, it seems natural  to
define  the  quantal  adiabatic  Berry  phase  for  open paths in
parameter space and then analyse it within the semiclassical  and
classical  limits. Towards the end, an application of the present
work will be pointed out , where  one  can  show  the  effect  of
open-path  Hannay  angle  on  wave-packet revivals. The effect of
Hannay angle on revivals has been recently discussed by Jarzynski
\cite{cj} for cyclic variations  of  external  parameters.  In  a
sense,  the  application  of  the  present formulation will be an
extension of his prediction which says that effect  of  adiabatic
variation  of  parameters  is  to  cause  a  displacement  of the
location at which the  revived  wave-packet  appears,  {\sl  even
though  the parameters do not return to their original value over
the revival time.} \\

\mbox{}\\
\centerline{{\bf 2. Adiabatic Berry phase for open paths}}\\
\mbox{}\\

\$ 2.a {\it Berry phase for closed-paths:}\\

\par
Before  providing  the  generalised  Berry  phase  formula, it is
useful to recapitulate the standard Berry phase formula. Consider
a quantum system which is bounded, integrable  and  driven  by  a
slowly changing Hamiltonian $H({\bf R}(t)), \{{\bf R} = R_i\}$ is
the  set  of  externally  controllable parameters. Then using the
adiabatic approximations, the  solution  to  the  Schr{\"o}dinger
equation is given by

\begin{equation}
|\Psi(t)> = exp[-{i\over \hbar} \int_0^t E_n(t)dt) exp(i{\gamma}_n(t))]|\Psi_n({\bf R}(t))>,
\end{equation}
where  $|\Psi_n({\bf  R}(t))>$'s are instantaneous eigenstates of
the Hamiltonian with non-degenerate eigenvalues $E_n(t)$'s.  This
foregoing  eq.(1)  says that the system remains in the eigenstate
with quantum number $n$ apart from phase factors. The first phase
factor is  the  usual  dynamical  one.  The  extra  phase  factor
$exp(i\gamma_n(t))$  becomes physically important and non-trivial
only when the parameters are changed along  a  closed  path  over
some  time (large enough) T, such that ${\bf R}(T) = {\bf R}(0)$.
Otherwise, these extra phases can always be chosen identically to
be zero by choosing a different  eigenfunction.  The  non-trivial
phase is the Berry phase for closed paths in the parameter space,
given by

\begin{equation}
\gamma_n(C) = i \oint_C <\Psi_n({\bf R})|\nabla \Psi_n({\bf R})>.d{\bf R}
            = \oint_C {\bf A}_n({\bf R}).d{\bf R}.
\end{equation}
This is nothing but the line integral of a vector potential ${\bf
A}_n({\bf R})$ (called Berry potential or Berry one-form) arround
the closed curve in parameter space and which can also be written
as  a  surface  integral  of  a vector field (two-form) where the
surface is bounded by the closed curve $C$.  As  is  well  known,
this is non-integrable in nature and depends only on the geometry
of  the  path  in the parameter space. In addition to this, it is
gauge invariant. The phase $\gamma_n(C)$ is  independent  of  the
rate  at  which  the  circuit  $C$  is  traversed,  provided  the
adiabatic approximation holds. Therefore, the Berry phase  in  an
essential  ingredient  of  the  adiabatic  cyclic  evolution of a
quantum system.

\$2.b {\it Generalisation of Berry phase to open-paths:}\\

\par
Suppose that the parameters which have been adiabatically changed
along  an  arbitrary  curve  $\Gamma$,  do not come back to their
original value after some time  $t_f$.  Can  we  still  assign  a
geometric  phase to such an adiabatically evolving quantum state?
The answer is yes, though the phase, in this case, is  not  given
by the expression

\begin{equation}
\gamma_n(\Gamma) = \int_{{\bf R}(0)}^{{\bf R}(t_f)} {\bf A}_n({\bf R}).d{\bf R}.
\end{equation}
In the past it has been mentioned incorrectly that the non-cyclic
Berry  phase  would  be  still  given  by  the  above  expression
\cite{wu}. The reason being that  the  above  expression  is  not
gauge   invariant   under  local  gauge  transformations  of  the
eigenstates. We call the  expression  of  the  type  (3)  as  the
``Berry  term'' and which reduces to the Berry phase for a closed
loop in parameter space. To obtain the Berry  phase  formula  for
open paths we have to take care of the contributions from the end
points of the open path. When we do that the whole expression can
be made gauge invariant.

\par
The  mathematical  and  physical  basis  underlying the open-path
Berry phase formula can be given in terms  of  the  fiber  bundle
descriptions   of  the  adiabatically  evolving  eigenstates.  As
illustrated by Simon \cite{bs}, the fiber bundle has a base space
$M$ (which is the space of parameters), has fibers  (the  set  of
phase factors, namely the group $U(1)$ ) and has the bundle space
$E$  (in which the adiabatic eigenstates exist). The bundle space
$E$ over $M$ is defined by associating ${\bf R} \rightarrow
|\Psi_n({\bf  R})>$  given  by  $H({\bf  R})|\Psi_n({\bf  R})>  =
E_n({\bf  R})|\Psi_n({\bf R})>$ with fibers $U(1)$. Geometrically
we can imagine that the  time  evolution  of  the  eigenstate  is
represented  by  a  path in the bundle space $E$. The path in the
bundle space can be constructed by the knowledge of path that  is
actually  followed  by  the  parameters  in  the  base space. For
example, in the  case  of cyclic change of parameters the path in
the base space is a closed curve, whereas the path in the  bundle
space  is an open one with the initial and final points belonging
to the same fiber. However, if the parameters do not come back to
their original value after some time $t_f$, then the  base  space
path is an open path and correspondingly the lift of this is also
an  open  path  in  the  bundle space. But in this later case the
initial and final points of the bundle path are not on  the  same
fiber.  And  we  are  concerned  here precisely with this type of
adiabatic evolutions.

\par
In  general   (irrespective   of   adiabaticity,   cyclicity  and
unitarity), when the initial and  final  state  of  the  evolving
quantal system belong to two different fibers, we can compare the
phases  by  taking the inner product between them. This is in the
spirit of Pancharatnam's \cite{sp}  way  of  defining  the  phase
difference  between  two  different polarisation states of light.
However, the only restriction here is that the initial and  final
states should not be orthogonal to each other. Let $|\Psi(t)> \in
{\cal  H}$  be  the  state  of  a system at some instant of time.
During a non-cyclic evolution of the state vector in  ${\cal  H}$
it  traces  an  open curve whose projection is also an open curve
$\Gamma: \rho(0) \rightarrow \rho(t)  \rightarrow  \rho(t_f)  \neq
\rho(0)  $  in  the  projective  Hilbert  space, where $\rho(t) =
|\Psi(t)><\Psi(t)|$ is pure  state  density  operator.  The  total
phase difference between the initial and final states is given by

\begin{equation}
\Phi_T = arg<\Psi(t_i)|\Psi(t_f)> =  arg<\Psi(0)|\Psi(t_f)>.
\end{equation}
Using the projective geometric structure of the Hilbert space, it
has been shown by the present author \cite{ak} that the geometric
phase during an arbitrary evolution of quantum system is given by
\begin{equation}
\Phi_g  = i \int <\chi(t)|d\chi(t)>,
\end{equation}
where  $|\chi(t)>$  is  a  ``reference-section'' defined from the
actual state as $|\chi(t)> = {<\Psi(t)|\Psi(0)> \over
|<\Psi(t)|\Psi(0)>|}|\Psi(t)>$ and $i<\chi(t)|d\chi(t)>$   is  a  connection-form
defined  over the projective Hilbert space of the quantum system.
Thus, $\Phi_g$ can be regarded as  the  holonomy  of  the  $U(1)$
bundle  over  the  projective  Hilbert  space  ${\cal  P}$ of the
quantum system.

When  the quantum evolution is necessarily adiabatic and the open
path arises from the adibatic evolution of external parameters,
then  we obtain the open-path Berry phase, which is given
by
\begin{equation}
\gamma_n(\Gamma) = i \int_{\Gamma} <\chi_n({\bf R})|\nabla \chi_n({\bf R})>.d{\bf R} = \int_{\Gamma} \Omega_n({\bf R}).d{\bf R},
\end{equation}
where  $|\chi_n(R)>$ is the ``reference-eigenstate'' defined from
the adiabatic eigenstate as
 $|\chi_n({\bf R})> = {<\Psi_n({\bf R})|\Psi_n({\bf R}(0))> \over
|<\Psi_n({\bf R})|\Psi_n({\bf R}(0))>|}|\Psi_n({\bf R})>$. This
can  be  obtained from (5) by inserting the adiabatic approximate
wavefunction as given  in  (1).  We  have  denoted  the  adibatic
open-path  Berry  phase as $\gamma_n(\Gamma)$ to distinguish from
more general geometric phase $\Phi_g$. Thus, the adiabatic  Berry
phase  is  nothing  but  the line integral of a generalised gauge
potential   $\Omega_n({\bf   R})   =   i<\chi_n({\bf   R})|\nabla
\chi_n(R)>$  over  the parameter space. The relation between this
gauge potential and Berry potential can  be  worked  out  and  it
follows that

\begin{equation}
\Omega_n({\bf R}) = {\bf A}_n({\bf R}) - {\bf P}_n({\bf R}),
\end{equation}
where ${\bf P}_n({\bf R})$ is a new gauge potential, given by
\begin{eqnarray}
{\bf P}_n({\bf R}) & = & {i \over 2 |<\Psi_n({\bf R}(0))|\Psi_n({\bf R})>|^2 }  \bigg[<\Psi_n({\bf R}(0))|(|\nabla\Psi_n({\bf R})><\Psi_n({\bf R})| -  \nonumber\\
                   &   & |\Psi_n({\bf R})><\nabla \Psi_n({\bf R})|)|\Psi_n({\bf R}(0))> \bigg].
\end{eqnarray}
By  virtue  of  its  transformation  property under a local gauge
transformation  one  can make sure that ${\bf P}_n({\bf R})$ is a
vector potential in the parameter space (see below).  Thus,  like
the Berry potential ${\bf A}_n({\bf R})$, ${\bf P}_n({\bf R})$ is
a  vector potential defined over the whole parameter space except
that the later depends on the initial point  of  the  curve.  For
example,  if  we  change  the  initial value of the parameter the
value of the gauge potential  will  be  different.  Infact,  this
property  of the gauge potential ${\bf P}_n({\bf R})$ ensures the
non-integrable nature of the open-path Berry phase.

Now  the  open-path  Berry  phase  can be given a {\it gauge theoretic
description} in terms of these potentials as

\begin{equation}
\gamma_n(\Gamma) = \int_{\bf R(0)}^{\bf R(t_f)} [{\bf A}_n({\bf R}) - {\bf P}_n({\bf R})].d{\bf R},
\end{equation}
which says that the open-path Berry phase is the line integral of
the  difference  of  these two potentials in the parameter space.

\par
This phase has the following properties. It is real, because both
the  potentials are real. It is independent of the parameter that
we use to parametrise the evolution curve.  It  does  not  depend
explicitly  on the Hamiltonian or eigenvalue of the system. It is
non-additive in nature which in turn attributes a memory  to  the
adiabatically  evolving  quantal state. Hence, it qualifies to be
called as the Berry phase for open-paths in parameter space.  One
can  check  that  in the limiting case, the open-path Berry phase
formula obtained by us, precisely goes over to the  cyclic  Berry
phase when the parameters come back to their original value after
some time $t_f = T$.

Next  we  explicitly  show  the invariance of the open-path Berry
pahse under gauge and phase transformations. Under  $U(1)$  local
gauge  transformation  of  the adiabatic eigenstate $|\Psi_n({\bf
R})>$, we have  $|\Psi_n({\bf  R})>  \rightarrow  e^{i\alpha({\bf
R})}  |\Psi_n({\bf  R})>$.  It induces a gauge transformations on
${\bf A}_n({\bf R})$ as well as on ${\bf P}_n({\bf R})$:

\begin{eqnarray}
{\bf A}_n({\bf R}) & \rightarrow & {\bf A}_n({\bf R}) - \nabla \alpha({\bf R}) \nonumber\\
{\bf P}_n({\bf R}) & \rightarrow & {\bf P}_n({\bf R}) - \nabla \alpha({\bf R})
\end{eqnarray}
Therefore,  the open-path Berry phase is clearly gauge invariant,
because under local gauge transformations these vector potentials
transform  in  the  same  way  and  hence  their  difference   is
gauge-compensated.

Further,  it  can be shown that the open-path Berry phase is also
invariant under phase transformations. On redefining  the  phases
of the adibatic eigenstate as

\begin{equation}
|\Psi_n({\bf  R})> \rightarrow
|\Psi_n({\bf  R})> exp(i\int_{0}^{{\bf R}} {\bf K}({\bf R'}).d{\bf R'}),
\end{equation}
we can see that it affects both the vector potentials. The Berry potential
and the new potential undergo transformations as
\begin{eqnarray}
{\bf A}_n({\bf R}) & \rightarrow & {\bf A}_n({\bf R}) - {\bf K}({\bf R}) \nonumber\\
{\bf P}_n({\bf R}) & \rightarrow & {\bf P}_n({\bf R}) - {\bf K}({\bf R})
\end{eqnarray}
Therefore,  the  open-path Berry phase is unchanged under a phase
transformation. These properties enables us to define the concept
of Berry phase {\it  even  for  an  infinitesimal  path}  in  the
parameter space. For example, if the parameters are changed by an
amount  $\Delta  {\bf  R}$, the corresponding change in the Berry
phase would be given by

\begin{equation}
\Delta  \gamma_n   =  [{\bf  A}_n({\bf  R})  -   {\bf   P}_n({\bf
R})].\Delta{\bf R}.
\end{equation}

Here, some remarks concerning the gauge potential ${\bf P}_n({\bf
R})$ can be made as to whether it is a new geometric structure on
the  Hilbert space of quantum states. We will show that it is not
only a {\it new geometric structure} but also can be regarded  as
a  much  {\it richer gauge structure} in the sense that the Berry
potential is only a part of it. Indeed, we will show that it  can
be  split into two parts: one is just the Berry potential and the
other is related to the matrix elements of product of  projection
operators  and  force  operator (force operaor is $-\nabla H({\bf
R})$). To see this explicitly, let  us  express  ${\bf  P}_n({\bf
R})$ as

\begin{equation}
{\bf P}_n({\bf R}) = {i \over 2} \bigg[{<\Psi_n({\bf R}(0))|\nabla\Psi_n({\bf R})> \over <\Psi_n({\bf R(0)})|\Psi_n({\bf R}>}
                   - {<\nabla \Psi_n({\bf R})|\Psi_n({\bf R}(0))> \over <\Psi_n({\bf R})|\Psi_n({\bf R}(0))>} \bigg].
\end{equation}

On  inserting  a  complete  set of eigenstates at parameter value
${\bf R}$, we have
\begin{equation}
{\bf  P}_n({\bf  R}) =
{\bf  A}_n({\bf  R})  -  Im
\sum_{m  \not=  n}  {<\Psi_n({\bf  R}(0))|\Psi_m({\bf  R})> \over
<\Psi_n({\bf R}(0))|\Psi_n({\bf  R})>}  {<\Psi_m({\bf  R})|\nabla
H|\Psi_n({\bf R})> \over
             (E_n({\bf R}) - E_m({\bf R}))}
\end{equation}
where,   we   have   used   the  fact  that  ${\bf   A}_n  =  -Im
<\Psi_n|\nabla \Psi_n>$.
The above expression clearly shows  the  richness  of  the new gauge
structure  and  brings  out  the fact that the Berry potential is
only a part of it. Also, it provides a suitable formula  for  the
open-path Berry phase as
\begin{equation}
\gamma_n({\Gamma}) =
\int_{\Gamma} Im
\sum_{m  \not=  n}  {<\Psi_n({\bf  R}(0))|\Psi_m({\bf  R})> \over
<\Psi_n({\bf R}(0))|\Psi_n({\bf  R})>}  {<\Psi_m({\bf  R})|\nabla
H|\Psi_n({\bf R})> \over
             (E_n({\bf R}) - E_m({\bf R}))}.d{\bf R},
\end{equation}
which clearly shows the independence of the choice of  the  phase
of the eigenstates. (One may recall the expression for the  field
strength  ${\bf V}_n$ which was provided in the original paper of
Berry \cite{mb} and note the similarity here.) The  formula  (16)
is  very  useful and has been recently studied in connection with
linear response  theory  of  adiabatic  quantum  systems  and  in
understanding  the  damping  of  collective  excitations in Fermi
systems \cite{sj}, where the  dynamics  is  chaotic.  Also,  this
generalised  Berry  phase  theory  has  been  applied to physical
systems (like collection  of  electrons  and  nuclei)  where  one
applies  Born-Openheimer  approximation  and it is found that the
quantum fluctuation in the generator of the parameter  change  is
related  to  the time correlation function of the ``fast'' system
\cite{ap}, thus establishing a fluctuation-correlation theorem in
many-body context. The  connection  between  the  quantum  metric
tensor, force-force correlation and the open-path Berry phase has
been discussed for integrable and chaotic quantum systems.

~~~~~~\\
\centerline{{\bf 3. Connection between Hannay angle  and  Berry
phase using Coherent states}}\\
\mbox{}\\

\par
Consider  the  classical  counter part of the quantum system with
$N$ degrees of freedom, where the Hamiltonian of  the  system  is
given by $H({\bf q},{\bf p},{\bf R})$. We assume that there exist
$N$  constants  of motion in involution and the dynamical system is thus
integrable.  The   classical   trajectories   are   confined   to
$N$-dimensional  manifold,  which is an $N$-dimensional torus. It
is  known  that  for  integrable  systems,  we  can  go  over  to
action-angle  $(I_i,  \theta_i), i =1,2,....N$, description where
the actions remain invariant during an adiabatic  excursion.  The
angle  variables undergo additional shift (Hannay angle) during a
cyclic variation of  parameters.  The  total  change  in  angular
coordinate of the trajectory in phase space is thus given by

\begin{equation}
\theta_i(T)   =  \theta_i(0)  +  \int_o^T  \omega_i({\bf  I},{\bf
R(t)})dt + \oint <{ \partial \theta_i \over \partial {\bf R}}>.d{\bf
R},
\end{equation}
The  above  expression  consists  of   a  dynamical  angle  shift
(given by time integral of the  instantaneous  frequency)  and  a
geometric  angle  shift,  the  later  being known as Hannay angle
\cite{jh}. Like Berry connection does not depend  on  the  precise
form  of  the  Hamiltonian  but only on its symmetries, similarly
Hannay one-form  depends  on  the  symmetries  of  the  classical
Hamiltonian.   The   symmetries   in   this  case  are  canonical
automorphism of the invariant tori in phase space \cite{gkm}. The
standard formula for Hannay angle, however, is not valid  if  the
parameters are not brought back to their original value. Because,
under a rotation with respect to the angle variables of the phase
space  trajectories  the  Hannay angle does not remain invariant.
Remembering the difficulties encountered in this  problem,  which
we  have mentioned in the introduction, it is natural to look for
the connection between the open-path Berry phase and  geometrical
angle shift.

Here,  we bring out the connection between the phase holonomy and
angle holonomy using the parametrised  coherent  state  formalism
that  describes  the  action and angle variables in the classical
limit. In the sequel, we closely follow the methods of  Maamache,
Provost  and  Vallee  \cite{pv}.  For simplicity, let us restrict
ourselves first to one degrees of freedom. Given an adiabatically
changing Hamiltonian $H({\bf R})$ we can define a coherent  state
for the quantum system as

\begin{equation}
|\alpha, {\bf R}> = e^{-|\alpha|^2 /2} \sum_{n=0}^{\infty} {\alpha^n \over \sqrt n!} |\Psi_n({\bf R})>.
\end{equation}
We  can  also  define  an excitation operator or quantum counting
operator $N({\bf R})$ as
\begin{equation}
N({\bf R}) = \sum_{n=0}^{\infty} n |\Psi_n({\bf R})><\Psi_n({\bf R})|
\end{equation}
and $N({\bf R})$ satisfies an eigenvalue equation
\begin{equation}
N({\bf R})|\Psi_n({\bf R})> = n |\Psi_n({\bf R})>
\end{equation}
In  the  classical  limit  ($\hbar  \rightarrow  0, n \rightarrow
\infty$) the action is related to the excitation number $n$ as $I
= n\hbar$, which is finite. The coherent state is best suited for
studying the classical limit as it  represents  a  point  in  the
phase  space. The evolution of the coherent state represents
the trajectory along which the actions remain invariant.  Quantum
mechanically,  $|\alpha|^2$  represents  the  average value of the
counting operator and in the  classical  limit  $\hbar|\alpha|^2$
represents  invariant  action.  Physically  it  has  been  argued
\cite{pv} that
the  complex  parameter  $\alpha(t)$ is related to the action and
angle variable of the system as
\begin{equation}
\alpha(t) = \sqrt{I \over \hbar} e^{-i\theta(t)}.
\end{equation}

We  can  also  express  the  adiabatic  eigenstate  in  terms  of
action-angle state using the over completeness  property  of  the
coherent state. Since
\begin{equation}
|\Psi_n({\bf R(t)})> =  {1 \over \pi} \int d^2 \alpha e^{-|\alpha|^2 /2} {\alpha*^n \over \sqrt n!} |\alpha, {\bf R}>,
\end{equation}
where   $d^2\alpha   =   d(Re  \alpha)d(Im  \alpha)  =  {1  \over
2\hbar}dId\theta$, we can express the correspondence between  the
qunatum  eigenstate  and  a  point in phase space parametrised by
action and angle variable as
\begin{equation}
|\Psi_n({\bf R(t)})> =  {1 \over 2\pi {\hbar}^{(n/2 + 1)}} \int dI~d\theta e^{-I^2 /2\hbar^2} I^{n/2}  {e^{-in\theta} \over \sqrt n!} |I,\theta,{\bf R}>,
\end{equation}
where we have denoted $|\alpha, {\bf R}> = |I,\theta,{\bf R}>$.

\par
As the system evolves from some parameter value ${\bf R}(0)$, the
classical trajectory starts from some initial angle coordinate on
constant  action  surface. We wish to compute what would be the angle
shift  for some  arbitrary  parameter   value   ${\bf   R}(t_f)$.
Quantally, consider the evolution of the initial coherent state
$|\alpha(0),{\bf R}(0)>$. Then, at a later time t, the state is
given by
\begin{eqnarray}
|\alpha(t), {\bf R}(t)> & = & U(t)|\alpha(0),  {\bf R}(0)> \nonumber\\
                        & = & e^{-|\alpha|^2 /2} \sum_{n=0}^{\infty} {\alpha^n \over \sqrt n!} e^{i\Phi_n(t)} |\Psi_n({\bf R}(t))>,
\end{eqnarray}
where we have used the fact that $U(t)|\Psi_n({\bf R}(0))> = e^{(i\delta_n(t) + i\gamma_n(t))}
|\Psi_n({\bf R}(t))> = e^{i\Phi_n(t)} |\Psi_n({\bf  R}(t))>$.  Since,  in  the classical limit ,the sum
over $n$ is highly peaked arround the  value  $N  =  |\alpha|^2$,
most of the contribution to the sum comes from $n = N$. With this
idea, we can expand $\Phi_n(t)$ to first order in $(n -N)$
\begin{equation}
\Phi_n(t) = \Phi_N(t) + (n - N){\partial \Phi_N(t) \over \partial
N}
\end{equation}
Now, the parametrised coherent state at a later time t  is  given
by
\begin{equation}
|\alpha(t), {\bf R}(t)>  =
e^{i(\Phi_N(t) - N{\partial \Phi_N(t) \over \partial N})} |\alpha(0)e^{i{\partial \Phi_N(t) \over \partial N}},{\bf R}(t)>.
\end{equation}

To know the angle shift during a non-cyclic variation of external
parameters, we take the inner product of the  initial  and  final
(at time $t = t_f$) coherent state, which is given by
\begin{eqnarray}
<\alpha(0), {\bf R}(0)|\alpha(t_f), {\bf R}(t_f)> & = & e^{i(\Phi_N(t_f)   -   N{\partial   \Phi_N(t_f)   \over  \partial  N}}) e^{-|\alpha|^2}
                                                        \sum_{n,m}{\alpha(0)*^n  \over \sqrt n!} {\alpha(0)^m \over \sqrt m!}
                                                          \nonumber\\
                                                  &   & e^{im{\partial \Phi_N(t_f) \over  \partial  N}}  <\Psi_n({\bf
R}(0))|\Psi_m({\bf R}(t_f))>
\end{eqnarray}
Using random phase approximation, one can neglect terms $n \neq m$
and thus the above expression reduces to
\begin{eqnarray}
<\alpha(0), {\bf R}(0)|\alpha(t_f), {\bf R}(t_f)> & = &  e^{i(\Phi_n(t_f) - N{\partial \Phi_N(t_f) \over \partial N}})
                                                         e^{-|\alpha|^2} \sum_{n} {|\alpha|^{2n} \over  n!} \nonumber\\
                                                  &   & e^{i n{\partial \Phi_N(t_f) \over \partial N}}
e^{i\beta_n(t_f)}
|<\Psi_n({\bf R}(0))|\Psi_n({\bf R}(t_f))>|,
\end{eqnarray}
where   $\beta_n(t_f)  =  \int_{{\bf  R}(0)}^{{\bf  R}(t_f)}~{\bf
P}_n({\bf R}).d{\bf R}$. Following a similar argument  as  above,
we replace the phase $\beta_n(t_f)$ in the classical limit to its
first order approximation, viz, $\beta_n(t_f) = \beta_N(t_f) + (n
-  N){\partial  \beta_N(t_f)  \over  \partial N}$. Therefore, the
inner product between the initial and  final  coherent  state  is
given by

\begin{eqnarray}
<\alpha(0), {\bf R}(0)|\alpha(t_f), {\bf R}(t_f)> & = & e^{i(\Phi_N(t_f) - N{\partial \Phi_N(t_f) \over  \partial  N})}
                                                        e^{i(\beta_N(t_f) - N{\partial \beta_N(t_f) \over \partial N})} \nonumber\\
                                                  &   &e^{-|\alpha|^2} \sum_{n} {|\alpha|^{2n} \over  n!}
e^{i n({\partial \Phi_N(t_f) \over \partial N} + {\partial \beta_N(t_f) \over \partial N})}
|<\Psi_n({\bf R}(0))|\Psi_n({\bf R}(t_f))>|,
\end{eqnarray}
The  phase  factors appearing out side the summation are just the
global phase factors and do not contribute to the relative  phase
shift  of  the adibatic eigenstate, which would correspond in the
classical limit to the relative angle shift. Therefore, the total
angle shift would be given by the terms that  appear  inside  the
summation, i.e

\begin{equation}
\theta(t_f) - \theta(0)  = \triangle \theta = -{\partial \Phi_N(t_f) \over \partial N} + {\partial \beta_N(t_f) \over \partial N}
                         =  -{\partial \delta_N(t_f) \over \partial N}  - {\partial \gamma_N(\Gamma) \over \partial N}
\end{equation}
where  the  first term is the usual dynamical angle shift and the
second term  ${\partial  \gamma_N(\Gamma)  \over  \partial  N}  =
{\partial \over {\partial N}}{(\gamma_N(t_f) - \beta_N(t_f))}$ is
the  geometrical  angle  shift  or  Hannay  angle  for  open-path
excursions of the parameters. Therefore, the  connection  between
the  Hannay angle and Berry phase in the classical limit is given
by

\begin{equation}
\theta(I, \Gamma) = - \hbar {\partial \gamma(I,\Gamma) \over \partial I}
\end{equation}

For  N-degrees of freedom, the system admits $I_i$ and $\theta_i,
i = 1,2,...N$ action and angle  variables,  respectively.  It  is
straightforward to generalise the connection between Hannay angle
and Berry phase using product coherent states $\Pi_i
|\alpha_i(t),{\bf R(t)}>$, where each $\alpha_i(t)$ describes
$I_i$th action and $\theta_i$ angle variable. When the parameters
follow a non-cylic variation, then each angle variable $\theta_i$
undergoes an additional shift given by

\begin{equation}
\theta_i({\bf I}, \Gamma) = - \hbar {\partial \gamma({\bf I},\Gamma) \over \partial I_i}.
\end{equation}\\

\mbox{}\\
\centerline{{\bf 4. Semiclassical limit and Hannay angle}}\\
\mbox{}\\

\par
In  the  foregoing  discussions  we  describe  how  to obtain the
semiclassical Berry phase and  the  Hannay  angle  for  open-path
excursions  in  parameter space. Berry \cite{mv} has analysed his
closed-path phase in the semiclassical limit  and  established  a
connection  to the classical Hannay angle. In the same spirit one
can analyse the open-path Berry phase and derive  the  expression
for   adiabatic   angle  holonomy  for  open-path  excursions  of
classical Hamiltonian.  In  the  semiclassical  analysis,  it  is
assumed that the eigenfunction is associated with a torus and the
actions  are  quantised  according  to  Bohr-Sommerfeld \cite{jb}
rule. The semiclasical expression for the  wavefunction  \cite{mv}
is

\begin{equation}
\Psi_n({\bf q};{\bf R}) = <{\bf q}|n({\bf R})> = \sum_{\alpha} a_{(\alpha)}({\bf q}, I;{\bf R})exp({i \over \hbar} S^{(\alpha)}({\bf q},I;{\bf R})
\end{equation}
where  the  amplitude  $a_{(\alpha)}^2  =  {1  \over  (2  \pi)^N}
{d\theta^{(\alpha)}  \over  d{\bf  q}}  =  {1  \over  (2  \pi)^N}
det({\partial  \theta_{i}^{(\alpha)}  \over  \partial  q_i})$ and
$\alpha$ labels different branches of the multivalued,  classical
generating  function  $S^{(\alpha)}({\bf q},I;{\bf R})$. Each of
the action $S^{(\alpha)}$ satisfy the  Hamilton-Jacobi  equation.
The  existence  of  an  invariant  Lagrangian  surface (torus) is
important on which the  multivalued  actions  $S^{(\alpha)}$  are
defined.  Using  this  wavefunction  it is interesting to get the
semiclassical Berry phase for open paths. Upon substitution,  one
will  have  two  terms  viz  the Berry term and the new term. The
Berry potential can be easily evaluated and is given by

\begin{equation}
{\bf A}_n({\bf R}) = - 1/ \hbar \int d{\bf q} \sum_{\alpha}{1 \over (2 \pi)^N} {d\theta^{(\alpha)} \over dq}
\nabla S^{(\alpha)}({\bf q},I;{\bf R})
\end{equation}
where  $\int  d{\bf  q}  = \prod_{j=1}^N \int_{-\infty}^{+\infty}
dq_j$ and in evaluating this, it  is  assumed  that  products  of
terms  from  different  branches  of  $\alpha$  do not contribute
because  they  give  rise  to  rapid  oscillations   and   cancel
semiclassically on integrating over q. The additional term is not
so  straight forward to evaluate. However, we provide the closest
simplified expression for it. Note that the the vector  potential
${\bf P}_n({\bf R})$ can be written as

\begin{equation}
{\bf P}_n({\bf R}) = - Im{<\Psi_n({\bf R}(0))|\nabla\Psi_n({\bf R}(t))> \over <\Psi_n({\bf R}(0))|\Psi_n({\bf R}(t))>}
\end{equation}
Within the semiclassical approximation this can be expressed as

\begin{equation}
{\bf P}_n({\bf R}) = {X(I;{\bf R}) \nabla Y(I;{\bf R}) - Y(I;{\bf R}) \nabla X(I;{\bf R}) \over (X(I;{\bf R})^2 + Y(I;{\bf R})^2)}
\end{equation}
where
\begin{equation}
X(I;{\bf R}) = \int d{\bf q} \sum_{\alpha} a_{(\alpha)}({\bf q}, I;{\bf R}(0)) a_{(\alpha)}({\bf q}, I;{\bf R})
\cos [{1 \over \hbar}(S^{(\alpha)}({\bf q},I;{\bf R})  -  S^{(\alpha)}({\bf q},I;{\bf R}(0))]
\end{equation}
and
\begin{equation}
Y(I;{\bf R}) = \int d{\bf q} \sum_{\alpha} a_{(\alpha)}({\bf q}, I;{\bf R}(0)) a_{(\alpha)}({\bf q}, I;{\bf R})
\sin [{1 \over \hbar}(S^{(\alpha)}({\bf q},I;{\bf R})  -  S^{(\alpha)}({\bf q},I;{\bf R}(0))]
\end{equation}

Here, also those terms in the above expression survive which come
from  the  product  of  the  same branches of $\alpha$. Thus, the
semiclassical Berry phase formula for the open path excursion  in
parameter space is given by

\begin{eqnarray}
\gamma_n(\Gamma) = - \int_{\bf R(0)}^{\bf R(t_f)} \bigg[{1 \over \hbar} \int d{\bf q} \sum_{\alpha}{1 \over (2 \pi)^N} {d\theta^{(\alpha)} \over dq} \nabla S^{(\alpha)}({\bf q},I;{\bf R}) \nonumber
\end{eqnarray}

\begin{equation}
+ {X(I;{\bf R}) \nabla Y(I;{\bf R}) - Y(I;{\bf R}) \nabla X(I;{\bf R}) \over (X(I;{\bf R})^2 + Y(I;{\bf R})^2)} \bigg].d{\bf R}
\end{equation}
In a simplified notation the above formula can be expressed as

\begin{equation}
\gamma_n(\Gamma) = - \int \bigg[{1 \over \hbar} < \nabla S^{(\alpha)}>
+ {X \nabla Y  - Y \nabla X \over (X^2 + Y^2)} \bigg].d{\bf R}
\end{equation}
where, the integral over ${\bf q}$ has been converted    to    an
integral over the angles using the Jacobians $a_{(\alpha)}^2$ and
the limits of the integration is supressed because the  curve  in
parameter  space is arbitrary. Unless otherwise stated, the above
limit is understood as starting from some initial value to  final
value  of  the parameters. This is the semiclassical limit of the
open-path Berry phase.

During an adiabatic transport arround a closed-circuit, the above
expression  reduces  to  that  of  the well known result of Berry
\cite{mv}. When the time $t_f$ is so choosen that ${\bf  R}(t_f)
=  {\bf  R}(T)  =  {\bf  R}(0)$,  then  the  last  term  does not
contribute  to  the  semiclassical geometric phase,   i.e.   the   closed
line-integral over the parameters gives us

\begin{equation}
\oint [{X \nabla Y  - Y \nabla X \over (X^2 + Y^2)}].d{\bf R} = 0, mod~~ 2\pi n
\end{equation}
and hence the closed-circuit Berry phase for a loop $C$ is given by

\begin{equation}
\gamma_n(C) = {\int \int}_{\partial A = C} {\bf V}_n({\bf R}).d{\bf S}
\end{equation}
with
${\bf V}_n({\bf R}) = {1 \over \hbar} \nabla \wedge \int < \nabla S^{(\alpha)}>.d{\bf R}$

\par
Next,  we  obtain  the  Hannay  angle  for adiabatically evolving
systems around an open circuit that  has  been  promised  in  the
begining  of this paper. Using the connection between the quantal
geometric phase and the classical Hannay angle  one  can  express
the later as

\begin{equation}
\theta_H(I;\Gamma) = - \hbar {\partial \gamma_n(\Gamma) \over \partial I}
\end{equation}
Therefore,  the  classical  angle  Holonomy  during the adiabatic
variation of the Hamiltonian along an arbitrary path in parameter
space connecting the points ${\bf R}(0)$ and  ${\bf  R}(t_f)$  is
given by

\begin{equation}
\triangle \theta_H(I;\Gamma) = {\partial \over \partial I} \int < \nabla S^{(\alpha)}>.d{\bf R}
+  {\hbar \over (X_f^2 + Y_f^2)} (X_f {\partial Y_f \over \partial I}  - Y_f {\partial X_f \over \partial I})
\end{equation}
where $X_f = X(I;{\bf R}(t_f))$ and $Y_f = Y(I;{\bf R}(t_f))$.

\par
Thus,  the  adiabatic  system  admits a Hannay angle for {\it for
an open-path} which is the semiclassical  limit  of  the  quantal
adiabatic  Phase. {\it The open-path Berry phase and its relation
to Hannay angle constitute the  central  results  of  these  last
sections.}  Original Hannay angle (for closed-paths) is invariant
under parameter-dependent and action-dependent transformations of
the origin from which the angle $\theta$'s  are  measured.  Here,
the  generalised  Hannay  angle  will remain invariant under more
general  type  of  parameter-depenedent  and   action   dependent
transformations. The additional term takes care of the invariance
of  the  Hannay angle under arbitrary transformations. At quantal
level, this property corresponds to the invariance  of  open-path
Berry  phase  under  parameter dependent phase transformations of
the eigenfunctions.\\

{\centerline {\bf 5. Classical limit of open-path Berry phase and
connection to Hannay angle}}

\vskip 1cm

\$ 5.a {\it Berry phase from instantaneous projectors}:\\

In  this  section, we intend to obtain the classical limit of the
Berry  phase  when  the  parameters  need  not  follow  a  cyclic
evolution.  Essentially,  the  problem  reduces  to  finding  the
classical limit of the generalised one-form $\Omega _n^{(1)}$  or
the  vector  potnetial  $\Omega  _n({\bf R})$, so that one may be
able to shed some light on what  would  be  the  classical  angle
holonomy  for  non-cyclic variations. To this end, we express the
open-path Berry phase in terms of the averages of the commutators
of  the  instantaneous  projection  operators  $P_n({\bf  R})   =
|\Psi_n({\bf  R})><\Psi_n({\bf  R})|$  as it will facillitate the
classical  limit  with  ease.  This  one-dimensional   projection
operator  depends  on  the  parameter  ${\bf}$  continuously  and
undergoes a continuous evolution in parameter space. Since we are
dealing  with  non-cyclic  evolutions  of  parameters   $P_n({\bf
R(t_f)})$ is not equal to $P_n({\bf R(0)})$. To express the Berry
phase in terms these projectors, note that (5) can be written as

\begin{equation}
\gamma_n(\Gamma) = {i \over 2}\int_{\Gamma}\bigg(<\chi_n({\bf R})|\nabla \chi_n({\bf R})> -
<\nabla \chi_n({\bf R})| \chi_n({\bf R})>\bigg).d{\bf R}
\end{equation}
By expressing the ``reference-eigenstate''$|\chi_n({\bf R})>$ as
$|\chi_n({\bf R})> = {P_n({\bf R})|\Psi_n({\bf R}(0))>  \over
|<\Psi_n({\bf R})|\Psi_n({\bf R}(0))>|}$, we have
\begin{equation}
<\chi_n({\bf R})|\nabla \chi_n({\bf R})> =
{<\Psi_n({\bf R}(0))|P_n({\bf R}) \nabla P_n({\bf R})|\Psi_n({\bf R}(0))>  \over
<\Psi_n({\bf  R})|P_n({\bf R})|\Psi_n({\bf R}(0))>} - {1 \over 2}
{<\Psi_n({\bf R}(0))|\nabla P_n({\bf R})|\Psi_n({\bf R}(0))>  \over
<\Psi_n({\bf  R})|P_n({\bf R})|\Psi_n({\bf R}(0))>}
\end{equation}
Inserting the above equation into the geometric phase formula, we
can write the open-path Berry phase in terms of the commutator of
the  projector  and its gradient over the space of parameters, as
is given by

\begin{equation}
\gamma_n(\Gamma) = {i \over 2} \int
{<\Psi_n({\bf R}(0))|[P_n({\bf R}), \nabla P_n({\bf R})]|\Psi_n({\bf R}(0))>  \over
<\Psi_n({\bf  R})|P_n({\bf R})|\Psi_n({\bf R}(0))>}.d{\bf R}.
\end{equation}

Thus, the generalised phase one-form would be given by
\begin{equation}
\Omega_n^{(1)} = {i \over 2}
{<\Psi_n({\bf R}(0))|[P_n({\bf R}),  dP_n({\bf R})]|\Psi_n({\bf R}(0))>  \over
<\Psi_n({\bf  R})|P_n({\bf R})|\Psi_n({\bf R}(0))>}
\end{equation}
where  $d$  is  the  exterior  derivative  with  respect  to  the
parameters. A similar formula has been derived by Mead  \cite{md}
for  the  case  of  cyclic  evolutions  and  by Wagh and Rakhecha
\cite{w} for non-cyclic  evolutions  in  the  projective  Hilbert
space  of  the quantum system after the present author introduced
the concept of ``reference-state''. It is interesting  to  remark
that   the   open-path   Berry   phase  has  its  origin  in  the
non-commutativity of the instantaneous projection  operator  with
its  exterior  derivative in the parameter space, which is purely
of quantum mechanical in nature. This expression is more suitable
to  study  the  classical  limit  beacause  there  is  a   direct
correspondence  between  the  quantum  mechanical  commutator  of
hermitian operators and the classical valued Poisson bracket.\\

\$ 5.b {\it Classical limit of Berry phase}:\\

To  analyse  the  classical limit of open-path Berry phase we use
Wigner-Weyl representation of quantal  expression  and  take  the
lowest  order term (in powers of $\hbar$) that will correspond to
the classical limit  of  the  former.  In  Wigner  representation
\cite{ep}   the   quantum   mechanical   operator   $\hat  O$  is
representated as a phase space function $O_W({\bf  q},{\bf  p})$,
where

\begin{equation}
O_W({\bf q},{\bf p}) = \int d^N{\bf y} <{\bf  q}+{\bf  y}/2|\hat
O|{\bf  q} - {\bf  y}/2> e^{-i{\bf p}.{\bf y} / \hbar}.
\end{equation}
The  Weyl  symbol of the operator reduces to the classical valued
function in the $\hbar \rightarrow 0$ limit. If we  choose  $\hat
O$  to  be  a  density  operator  ${\hat  {\rho}} = |\Psi><\Psi|$
constructed from a pure state wave  function,  then  we  get  the
Wigner function

\begin{equation}
\rho_W({\bf q},{\bf p}) = \int d^N{\bf y} \rho({\bf  q} - {\bf  y}/2,
{\bf  q} + {\bf  y}/2)e^{-i{\bf p}.{\bf y} / \hbar}.
\end{equation}
Wigner  representation  of  phase  space  density and phase space
function is an alternate approach to ordinary  quantum  mechanics
where  one can talk of classical limit of various quantities with
ease. In this representation, we can express the average  of  the
commutator  as  a  phase  space average of the Weyl symbol of the
commutator between the projection operators, i.e.

\begin{equation}
<\Psi_n({\bf R}(0))|[P_n({\bf R}),  \nabla P_n({\bf R})]|\Psi_n({\bf R}(0))> =
\int d^N{\bf q} d^N{\bf p}~ P_n({\bf q},{\bf p})
([P_n({\bf R}),  \nabla P_n({\bf R})])_W({\bf q},{\bf p})
\end{equation}
and similarly, we have for the denominator
\begin{equation}
<\Psi_n({\bf R}(0))|P_n({\bf R})|\Psi_n({\bf R}(0))> =
\int d^N{\bf q} d^N{\bf p} P_n({\bf q},{\bf p}) P_n({\bf q},{\bf p},{\bf R}).
\end{equation}
The  Weyl  symbol  of  the  commutator is given in terms of Moyal
bracket
\begin{equation}
([P_n({\bf R}),  \nabla P_n({\bf R})])_W =
{2 \over i} P_n({\bf q},{\bf p},{\bf R}) \sin \sigma \nabla P_n({\bf q},{\bf p}, {\bf R}),
\end{equation}
where $\sigma$ is given by
\begin{equation}
\sigma = \sum_{i=1}^N {\hbar \over 2} \bigg( {{\partial}^{\leftarrow} \over \partial
{\bf p}}.{{\partial}^{\rightarrow} \over \partial {\bf q}}
 - {{\partial}^{\leftarrow} \over \partial
{\bf  q}}.{{\partial}^{\rightarrow}  \over  \partial  {\bf  p}},
\bigg).
\end{equation}
where  the  left  and  right  arrow on the differential operators
imply that they act on the functions which lie to  the  left  and
right,   respectively.  Since  we  are  interested  only  in  the
classical limit of the generalised  vector  potential,  the  Weyl
symbol  of the commutator goes over to the poisson bracket of the
corresponding distribution functions on phase  space.  Hence,  we
have

\begin{equation}
([P_n({\bf R}),  \nabla P_n({\bf R})])_W  \rightarrow
{1 \over i} \{P_n({\bf q},{\bf p},{\bf R}), \nabla P_n({\bf q},{\bf p},{\bf R})\}_{P.B}.
\end{equation}
Also,  for  an  integrable  system  we  know  that  the invariant
manifold is torus on which $N$ actions remains constant  and  the
initial phase space distribution can be taken as a microcanonical
distribution, where $P({\bf q},{\bf p})$ is given by \cite{mvb}

\begin{equation}
P_n({\bf q},{\bf p}) = {1 \over (2\pi)^N}~~ \delta^N({\bf I}({\bf q},{\bf p}) - {\bf I}),
\end{equation}
This  $N$-dimensional  delta  function  tells  us that the Wigner
function for an eigenstate is concentrated in the region  that  a
classical  orbit  visits  over  an infinite time. The phase space
average of any function is defined as

\begin{equation}
<f>_I = {1 \over (2\pi)^N} \int d^N{\bf q} d^N{\bf p}~~ f({\bf q},{\bf p},{\bf R})
~\delta^N({\bf I}({\bf q},{\bf p}) - {\bf I}).
\end{equation}
Therefore, the classical limit  of  the  the  generalised  vector
potential is given by
\begin{equation}
\Omega_c({\bf R}) =
{-{1 \over 2} \int d^N{\bf q} d^N{\bf p}~~ \delta^N({\bf I}({\bf q},{\bf p}) - I).
\{P({\bf q},{\bf p},{\bf R}),  \nabla P({\bf q},{\bf p}, {\bf R})\}_{P.B} \over
 \int d^N{\bf q} d^N{\bf p}~~ \delta^N({\bf I}({\bf q},{\bf p}) - {\bf I})
P({\bf q},{\bf p},{\bf R}) }.
\end{equation}
Thus, the classical angle holonomy  $\theta_H^c$ for integrable system would be given by
\begin{equation}
\theta_H^c = \int \Omega_c({\bf R}).d{\bf R} =
-{1 \over 2} \int
{ < \{P({\bf q},{\bf p},{\bf R}),  \nabla P({\bf q},{\bf p},{\bf R})\}_{P.B} >_I \over
< P({\bf q},{\bf p},{\bf R}) >_I }.d{\bf R},
\end{equation}
which suggests that the origin of the angle holonomy could be due
to  the  non-vanishing  nature  of the torus average of the phase
space density with its gradient in parameter space.  However,  it
is  not  at  all  clear to the author how to prove this statement
purely using classical arguments.\\

\$ 5.c. {\it Operational definition of Hannay angle for open-paths}:\\

Although it is difficult to derive the non-cyclic Hannay angle at
classical  level, we can try to give an operational definition of
it. This would  require  the  knowledge  of  the  {\it  classical
analog}  of  the  quantum  mechanical  inner  product  of any two
vectors in the Hilbert space of the quantum  system.  In  quantum
theory  the most important thing is the inner product between two
non-orthogonal states which  is  in  general  a  complex  number.
Physically, this represents the survival amplitude of a system in
a  certain state once it is prepared in a given initial state. Is
there any such thing in the classical world? This is  a  question
which  bothers  some  physicist that I know and the answer is not
quite clear. However,  we  can  try  to  see  what  is  the  {\it
classical  limit}  of quantum mechanical inner product. It may be
remarked that the square of the  modulus  of  the  inner  product
(transition  probability)  between two states can be expressed in
terms of Wigner functions and in the classical  limit  this  will
represent  the  overlap  integral of microcanonical distributions
corresponding to two possible configurations.

Consider  two quantum states $|\Psi_1> = |\Psi(0)>$ and $|\Psi_2>
= |\Psi(t)>$ whose inner product is defined on the Hilbert space of
the  quantum  system.  If  $U(t)$  is  the  unitary operator that
generates $|\Psi_2>$ from $|\Psi_1>$, then the inner product  can
be expressed as

\begin{equation}
<\Psi_1|\Psi_2> = <\Psi(0)|U(t)|\Psi(0)> = tr(\rho(0) U(t)) =  \int d^N{\bf q} d^N{\bf p}~~\rho_W({\bf q},{\bf p}) U_W({\bf q},{\bf p},t),
\end{equation}
which  is  nothing  but  the  phase  space average of the unitary
operator over Wigner distribution. The classical  limit  of  this
would correspond to the phase space average of classical function
that   generates  the  canonical  transformation.  For  adiabatic
eigenstates let $U({\bf  R}(t_f)),{\bf  R}(0))$  be  the  unitary
operator  that  relates  the  states $|\Psi_n({\bf R}(t_f))>$ and
$|\Psi_n({\bf  R}(0))>$.  Then  the  inner  product  between  the
initial  and  final  adibatic  eigenstatea  can  be written as an
average of the unitary operator  $U({\bf  R}(t_f)),{\bf  R}(0))$.
Thus,  $<\Psi_n({\bf  R}(0))|\Psi_n({\bf R}(t_f))> = <\Psi_n({\bf
R}(t_f))|U({\bf  R}(t_f)),{\bf   R}(0))|\Psi_n({\bf   R}(t_f))>$.
Since  any unitary operator can be written as $U = C + iS$, where
$C$ and  $S$  are  commuting  hermitian  operators,  the  quantum
mechanical  inner  product  is given by $<C> + i<S>$, which is in
general a complex  number.  We  replace  the  quantum  mechanical
averages by its classical ones, where the averages of $C$ and $S$
are taken over microcanonical distributions and are given by

\begin{eqnarray}
<C>_{\bf I}   =  {1\over  (2\pi)^N}  \int d^N{\bf q} d^N{\bf p}  C({\bf q},{\bf p},{\bf R}(t_f),{\bf  R}(0))
~\delta^N({\bf I}({\bf q},{\bf p}) - {\bf I}), \nonumber\\
\end{eqnarray}

\begin{equation}
<S>_{\bf I}   =  {1\over  (2\pi)^N}  \int d^N{\bf q}  d^N{\bf  p}
S({\bf q},{\bf p},{\bf R}(t_f),{\bf  R}(0))
~\delta^N({\bf I}({\bf q},{\bf p}) - {\bf I})
\end{equation}
Here,  as  before  the  averages  \cite{jh}  are taken arround the
Hamiltonian contour on which the point $({\bf q},{\bf  p})$  lies
and  they are functions of the action ${\bf I}$, intial and final
parameter  value.  The  quantities  $C_c({\bf   q},{\bf   p},{\bf
R}(t_f),{\bf  R}(0))$  and $S_c({\bf q},{\bf p},{\bf R}(t_f),{\bf
R}(0))$ are classical  valued  functions  whose  Poisson  bracket
vanishes  and  is  related  to  the  generator  of the canonoical
transformation in classical phase  space.  Therefore,  one  could
write  the  classical  analogue  of  the quantum mechanical inner
product as $<C>_{\bf I} + i<S>_{\bf I}$. With this idea  one  can
give an operational definition of the non-cyclic Hannay angle as

\begin{equation}
\theta_H(I;\Gamma) =  \int < {\partial \theta \over \partial
{\bf R}}>.d{\bf R}
+  tan^{-1} ({<S>_{\bf I} \over <C>_{\bf I}})
\end{equation}
where the first term  is  the  usual  Hannay term and second term
represents an additional angle coming from the  argument  of  the
classical  limit  of  the  quantum  mechanical  inner product. In
future one may be able  to  derive  the  open-path  Hannay  angle
within  the  classical  mechanics  -which seems to be a difficult
task at present.\\

\mbox{}\\
\centerline{{\bf 6. Discussion and Conclusion}}\\
\mbox{}\\

\par
In  this  section  we  discuss  briefly  an  application  of  the
open-path  Berry  phase  and conclude the formalism that has been
developed in this  paper.  The  open-path  Berry  phase  and  its
classical  counter  part  can have important applications in many
areas of physics. Here, we will illustrate how it shows up in  an
interesting way for the case of wave-packet revivals. The revival
phenomena refers to the case, where a quantal wave-packet spreads
following  a  classical  trajectory,  reassembles after some time
$T_R$ (called revival time), and then takes  the  course  of  the
classical  trajectory.  This  phenomena  \cite{ps} which was well
studied for  time-independent  Hamiltonians,  recently  has  been
generalised  by  Jarzynski \cite{cj} to the case of adiabatically
changing Hamiltonian systems. He has shown that if initially  the
quantal  wave-packet  is  at  some  point (say) $({\bf q}_0, {\bf
p}_0)$ in phase space, then the effect of  adiabatic  changes  of
external  parameters  can  be manifested as a displacement of the
location  of  the  revived  wave-packet   along   its   classical
trajectory. The amount by which the packet is shifted is equal to
the  adiabatic,  closed-circuit  Hanny angle. In carrying out his
analysis it is assumed that the external parameters are varied in
a cyclic manner and the time period T over which  the  parameters
return  to their original value is just equal to the revival time
$T_R$. He has concluded that the effect of Berry phase on revival
phenomena  is  meaningful  only  when  the  revival  time   $T_R$
coincides  with  the  cyclic time. As we have shown in this paper
the Berry phase and Hannay angle are not only  well  defined  for
closed paths but also for open paths, it must be now evident that
{\sl  effect  of classical Berry phase on wave-packet revival can
be seen even when the  parameters  do  not  come  back  to  their
original  value  at  time  $T_R$}.  Hence, we argue that the nice
conclusion of Jarzynski need  not  be  restrictive  to  the  case
considered  by him, although his analysis may need a modification
(to properly take into account the contributions coming from  the
new  vector  potential).  If  one  probes  the  location  of  two
identically prepared wave-packet during its evolution  along  the
classical  trajectory  by  keeping  the  parameters of one packet
constant  and  varying  the  parameters  of  the  other  in   any
desireable  way, one will be able to demonstrate the existence of
open-path Hannay angle in wave-packet revivals. By observing  the
relative  shift  in  the  locations of the revived packet one may
infer the effect of open-path Hannay angle.

To conclude this paper, in section-2, we obtained the Berry phase
for  quantum (whose classical counter part is integrable) systems
when  parameters  follow  an  open  path  during   an   adiabatic
evolution.  The  reason  for  such  a motivation has been clearly
brought out. It is  found  that  a  generalised  gauge  potential
(quantum  one-form) can be defined over the parameter space whose
line integral gives the Berry phase for open paths excursions  of
the  parameters.  The  open-path Berry phase is shown to be gauge
invariant and also phase invariant. Further, the non-cyclic Berry
phase goes over to the usual Berry phase formula for cyclic path.

The  classical angle Holonomy for open path is not know and there
is  no  way  to  proceed  because  for  non-cyclic  variations of
external parameters it is not clear  how  to  compare  the  angle
variables.  In  section-3,  We have provided a connection between
the open-path Berry phase and  Hannay  angle  using  parametrised
coherent    states,   that   describes   action-angle   variables
appropriately. It is found that the open-path Hannay angle can be
obtained by taking a partial derivative of  the  open-path  Berry
phase  with  respect  to  the  quantum  number in large $n$ limit
(classical limit).

In  section-4, using the semiclassical approximation for the wave
function   we  have  evaluated  the  open-path  Berry  phase  and
subsequently  derived  the  semi-classical  Hannay   angle.   The
open-path  Hannay  angle contains an extra term which is ususally
absent  for  cyclic  angle  holonomy  of  integrable  system.  In
section-5,  we  analysed  the  classical  limit  of  the  quantum
one-form by expressing it in  terms  of  the  commutator  of  the
instantaneous  projection operators with its exterior derivative.
This  enables  us  to  take  the   classical   limit   by   using
corresondence   rule  between  the  commutator  and  the  Poisson
bracket. Using Wigner representation of distribution function and
its classical counterpart we  expressed  the  angle  holonomy  in
terms  of  the torus averages of the Poisson bracket of the phase
space density with its exterior derivative. It may be argued that
the quantum mechanical inner product has a classical limit  which
gives  rise  to  an additional term in Hannay angle for open path
excursions. The operational definition of the  non-cyclic  Hannay
angle   is   given  within  the  classical  mechanics-{\it  whose
derivation is still an open problem}. As an application  we  have
outlined  how  this  angle  holonomy can have important effect in
wave-packet revivals. The future challenge lies  in  establishing
the  open-path Hannay angle purely from classical considerations.
Since not much is known about  this  interesting  angle  holonomy
when the parameters do not follow a closed path, it is hoped that
this work will be an important step in this direction.

\vskip 2cm
{\bf  Acknowledgements}: ~~~ It is a pleasure to thank Dr. A. G. Wagh,
Prof. R. Simon and N. Mukunda for useful discussions.

\renewcommand{\baselinestretch}{1}

\end{document}